# ARTICLE

## Fully printed flexible perovskite solar modules with improved energy alignment by tin oxide surface modification

Lirong Dong,[a] Shudi Qiu,[a] José Garcia Cerrillo,[a] Michael Wagner,[b] Olga Kasian,[a,b,c] Sarmad Feroze,[a] Dongju Jang,[a] Chaohui Li,[a] Vincent M. Le Corre,[a,b] Kaicheng Zhang,[a] Heiko Peisert,[d] Felix U Kosasih,[e] Charline Arrive,[f] Tian Du, *[a,b] Fu Yang,[a,g] Christoph J. Brabec, *[a,b] Hans-Joachim Egelhaaf[a,b]

Fully printed flexible perovskite solar cells (f-PSCs) show great potential for the commercialization of perovskite photovoltaics owing to their compatibility with high-throughput roll-to-roll (R2R) production. However, the challenge remains in the deficiency in controlling interfacial recombination losses of the functional layer, causing remarkable loss of power conversion efficiency (PCE) in industrial production. Here, a fullerene-substituted alkylphosphonic acid dipole layer is introduced between the R2R-printed tin oxide electron transport layer and the perovskite active layer to reduce the energetic barrier and to suppress surface recombination at the buried interface. The resulting f-PSCs exhibit a PCE of 17.0% with negligible hysteresis, retain 95% of their initial PCE over 3000 bending cycles and achieve a T95 lifetime of 1200 h under 1 sun and 65℃ in nitrogen atmosphere. Moreover, the fully printed flexible perovskite solar mini-modules (f-PSMs) with a 20.25 cm$^2$ aperture area achieve a PCE of 11.6%. The encapsulated f-PSMs retain 90% of their initial PCE after 500 h damp-heat testing at 65℃ and 85% relative humidity (ISOS-D3). This work marks an important progress toward the realization of efficient and stable flexible perovskite photovoltaics for commercialization.

## Introduction

Flexible perovskite solar cells (f-PSCs) show considerable commercialization potential due to their low cost and high power conversion efficiency (PCE).[1-3] Consequently, an increasing number of researchers are focusing on the transition from small-area f-PSCs to large-area flexible perovskite solar modules (f-PSMs) with the aim of accelerating the commercialization of perovskite photovoltaics (PV).[4-7] The relevance of this field is high-lighted by the recent report on the first fully roll-to-roll (R2R) coated perovskites modules.[8] Nevertheless, reports on fully printed f-PSCs, including printed electrodes, are still scarce up to date. Among all available coating and printing techniques, slot-die coating has received the most attention for upscaling perovskite PV.[9, 10] Slot-die coating has revolutionized manufacturing through R2R processing, enabling precise, high-throughput film deposition on flexible substrates, thereby boosting production rates and yield. Nonetheless, realizing the full potential of this method necessitates the capability of depositing all device layers via R2R coating from the liquid phase at sufficiently high rates. Hence, conventional metal electrodes such as those made by evaporating silver, gold, copper, and other expensive materials are not optimal for large-scale fabrication. Instead, low-cost carbon black has emerged as a promising alternative due to its

a. Institute of Materials for Electronics and Energy Technology (i-MEET), Friedrich-Alexander-Universität Erlangen-Nürnberg, Martensstraße 7, 91058 Erlangen, Germany. Email: tian.du@fau.de; christoph.brabec@fau.de; lirong.dong@fau.de.
b. Helmholtz Institute Erlangen-Nürnberg for Renewable Energy (HI ERN), Cauerstraße 1, 91058 Erlangen, Germany.
c. Helmholtz Zentrum Berlin GmbH, Hahn-Meitner-Platz 1, 14109 Berlin, Germany.
d. Institut für Physikalische und Theoretische Chemie, Universität Tübingen, Auf der Morgenstelle 18, Tübingen, Germany.
e. Department of Materials Science and Metallurgy, University of Cambridge, 27 Charles Babbage Road, Cambridge CB3 0FS, United Kingdom.
f. ARMOR, Siège social 20 rue Chevreul, 44105 Nantes, France.
g. College of Chemistry, Chemical Engineering and Materials Science, Soochow University, 215123, China.
Electronic Supplementary Information (ESI) available: [details of any supplementary information available should be included here]. See DOI: 10.1039/x0xx00000x

exceptional printability and stability. Similarly, rather expensive hole extraction materials (HTL) with narrow processing windows and poor stability are preferred for the preparation of small cells with record efficiencies but are less appropriate for large-scale processing.[11] This shift in preference is motivated by the need for cost-effective solutions that can efficiently cover large areas while maintaining a high level of performance and reliability. Recently, the straightforward and stable device architecture of n-i-p (n-type - intrinsic - p-type) stack with a carbon top electrode has ignited significant research interest by using scalable techniques like slot-die coating and its lab analogue based on $SnO_2$ exhibit serious interfacial charge recombination and intrinsic hysteresis behaviour. Previous reports have shown that achieving reduced interface recombination and hysteresis-free devices with $SnO_2$-based ETLs requires interfacial modification at the surface.[19-21]

The modification of $SnO_2$ surfaces by dip coating and spin coating fullerene-substituted alkylphosphonic acids (FAPA) molecules has been reported by our group previously.[22, 23] However, dip coating and spin coating are not viable for large-area R2R coating. In this work, we have transferred this concept to a scalable process, namely doctor blading. By systematic doctor blading for its manufacturing.[12-14] One of the primary challenges in developing highly efficient flexible n-i-p devices is the realization of a high-quality electron transport layer (ETL), which can facilitate efficient charge extraction and provide a uniform substrate with appropriate surface energy for the growth of perovskite crystals. Generally, the most frequently used materials for the ETL in n-i-p structures are metal oxides like titanium oxide ($TiO_2$), zinc oxide (ZnO) and tin oxide ($SnO_2$).[15-17] Among them, due to the unique characteristics of low-temperature processing and high electron mobility, $SnO_2$ has emerged as the most promising alternative for large area processing, particularly with R2R techniques.[18] However, PSCs variation of the molecular structure of the FAPA molecules, we have modified the $SnO_2$ layer by two fullerene derivatives that have strong interface dipole properties. This results in significant performance enhancement, due to improved energy level alignment, reduced interface recombination and suppressed ion migration at the ETL/perovskite interface.

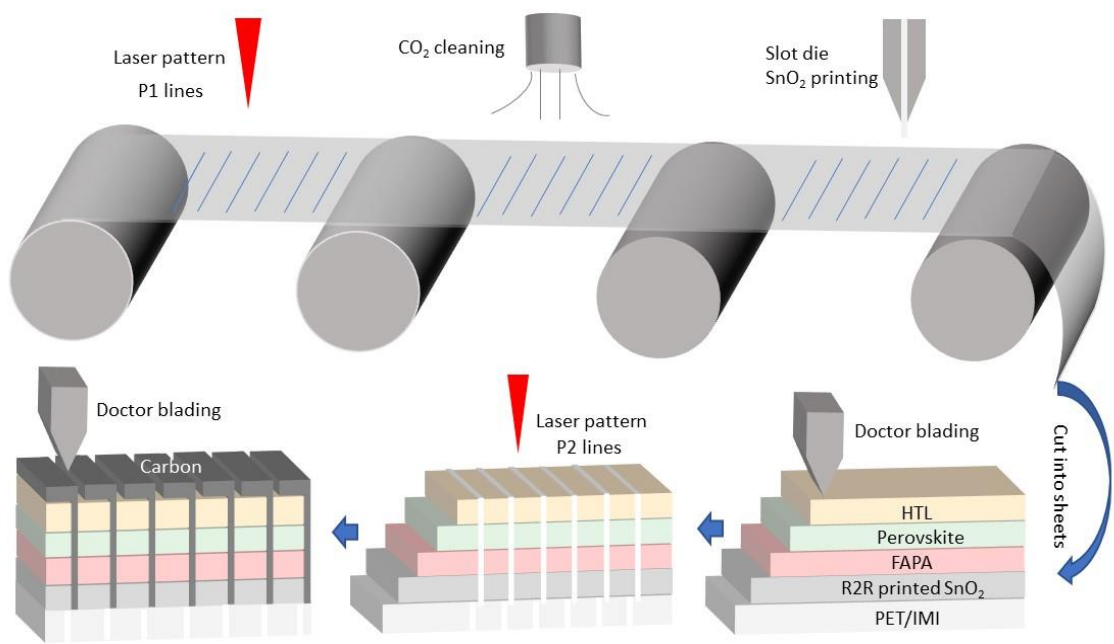

Figure 1 The workflow of manufacturing flexible perovskite solar modules, comprising laser patterning of the IMI bottom electrode, deburring by $CO_2$ snow jet cleaning, R2R slot-die coating of the ETL, deposition of FAPA interface dipole layer, perovskite active layer and HTL by doctor blading, laser patterning of the active layer stack (P2), and deposition of the carbon electrode by stencil printing (P3). Manufacturing of perovskite solar cells follows the same workflow, but without the laser patterning steps P1-P3 (for details see the experimental part in the SI).

Employing the optimized ETL/perovskite interface and a metal-free printed carbon top electrode, we demonstrate fully printed f-PSCs with a PCE of 17.0%. Besides, fully printed f-PSMs with 20.25 cm² aperture area (active area of 16.84 cm² and dead area of 3.41 cm²) are fabricated by doctor blading in ambient atmosphere, yielding a PCE of 11.6%. The encapsulated f-PSMs retain 90% of their initial PCE after 500 h damp-heat test at 65°C and 85% relative humidity (RH), conforming to the ISOS-D3 stability test protocol, which validates the potential towards the commercialization of printable flexible PV.

## Results

### Workflow of flexible solar cell and solar module fabrication

The f-PSCs and f-PSMs, employing the layer stack PET/IMI/SnO$_2$/FAPA/perovskite/HTL/Carbon, are prepared on alkylphosphonic acids (FAPA). As shown in **Figure 2b**, two different fullerene alkyl phosphonic acids, mono(alkylphosphonic acid)fullerene (referred to as "monoFAPA") and bis(alkylphosphonic acid)fullerene (referred to as "bisFAPA"), are investigated.

Representative current density-voltage (J-V) curves of the optimized FAPA layers are plotted in **Figure 2c**, as compared to a reference device comprising unmodified SnO$_2$. Modification of SnO$_2$ by monoFAPA improves device performance, mainly

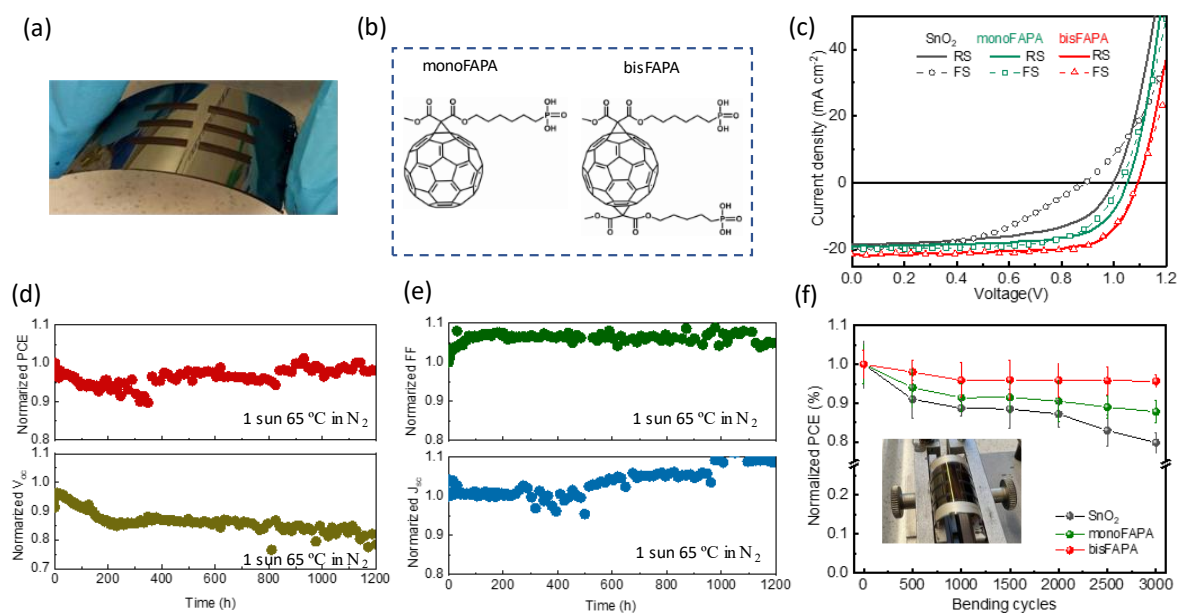

Figure 2 (a) Photograph of f-PSCs. (b) Chemical structures of monoFAPA and bisFAPA. (c) Representative J-V curves measured with forward scanning (FS) and reverse scanning (RS) for reference and FAPA-modified devices. (d)-(e) Illumination stability under 1 sun and 65°C in nitrogen atmosphere with bisFAPA modified device. (f) Normalized PCE of f-PSCs as a function of the number of mechanical bending cycles. The insert figure shows a photograph of the bending test.

polyethylene terephthalate (PET)/ITO-Silver-ITO (IMI) flexible substrates. The schematic drawing shown in **Figure 1** illustrates the workflow of device fabrication: The PET/IMI substrate is first patterned by laser ablation to separate the bottom electrodes of the individual cells (P1). SnO$_2$ is coated onto the patterned substrate using a slot-die R2R technique. The foil of the SnO$_2$-coated substrate is then cut into small pieces for the coating of other layers by doctor blading, including perovskite, the HTL, and the top carbon electrode. This entire process takes place in ambient air and at low temperatures (≤ 120 °C).

### Impact of SnO$_2$ surface modification on solar cell performance

We first consider reducing ETL/perovskite interfacial recombination in small-area f-PSCs (**Figure 2a**), through modification of the SnO$_2$ surface with a thin layer of fullerene contributed by an increase in average fill factor (FF) from 53% to 64% and an improvement of the average $V_{OC}$ by 50 mV, whilst modification by bisFAPA leads to a greater improvement of FF to 69% and of $V_{OC}$ by 100 mV with respect to the reference (for detailed parameters see **Figures S1-2**). Obviously, bisFAPA outperforms monoFAPA in terms of modifying the SnO$_2$ surface. The champion cell thus achieved by using bisFAPA shows a short-circuit current density ($J_{SC}$) of 21.7 mA cm$^{-2}$, $V_{OC}$ of 1.09 V, FF of 72% and PCE of 17.0%. The $J_{SC}$ measured from J-V scan is in acceptable agreement with the integrated $J_{SC}$ from external quantum efficiency (EQE) measurements, which show values of around 20.0 mA cm$^{-2}$ (**Figure S3**).

Notably, as shown in **Figure 2c**, whilst the reference device exhibits significant J-V hysteresis, it is reduced for the monoFAPA modified devices and is almost completely absent for the bisFAPA modified devices. The observed phenomenon is likely a result of the diminished density of trap defects within the perovskite with high-quality.[24] The champion cell with bisFAPA shows steady-state power output (SPO) of 16.7% through 100 seconds photocurrent measurement at the maximum power point, as demonstrated in **Figure S4**, higher than that of the monoFAPA modified device (13.1%) and the reference device (10.0%). To our knowledge, this is the highest value ever reported for fully printed f-PSCs.

unencapsulated solar cells under 1 sun illumination, **Figure 2d-e**, show impressive long-term stability upon modification by bisFAPA, with around 95% of the initial efficiency remaining after 1200 h aging. In comparison, monoFAPA-modified devices exhibit 85% retention, while the reference devices show only 55% of their initial PCE (**Figure S6**). Evolution of PCE during bending test of the f-PSCs, **Figure 2f**, highlights improved mechanical stability of bisFAPA-modified interfaces, after 3000 bending cycles with a bending radius of 5 mm, the devices maintain 95% of their initial PCE. In comparison, monoFAPA-

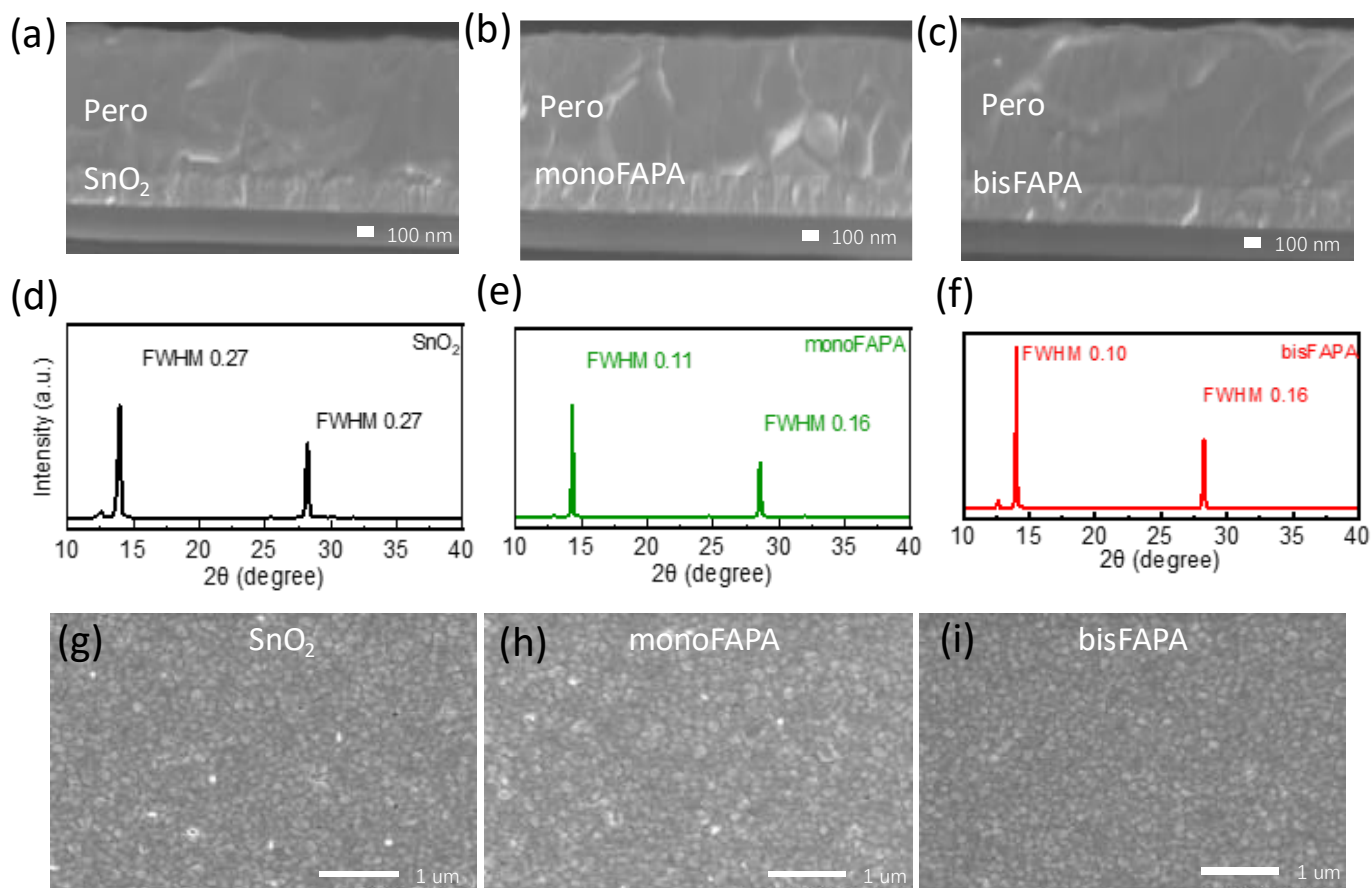

Figure 3 (a)-(c) Cross-sectional SEM of perovskite on SnO₂, monoFAPA and bisFAPA modified SnO₂ interfaces. (d)-(f) the XRD pattern of perovskite on bare SnO2 and two FAPA modified interfaces. (g)-(i) Top-view of perovskite on different interfaces after 3000 bending circles.

Moreover, a comparison has been conducted between metal electrode-based devices and those utilizing carbon electrodes, revealing negligible performance loss (**Figure S5**). Notably, the stability is significantly enhanced when employing stable carbon electrodes, as demonstrated in our prior work.[12-14]

Before investigating the origin of performance enhancement by FAPA modification, we first assess the illumination stability and mechanical stability. The operational lifetime tests of modified devices retain 88% of their initial PCE, while the reference devices show a lower retention of only 80%. The original average PCEs are depicted in **Figure S7**. This underscores the significantly superior mechanical stability of bisFAPA-modified interfaces when compared to both monoFAPA and the reference devices.

**Crystallization of perovskite on FAPA modified interfaces**

In order to elucidate the impact of interfacial modification by FAPA on the improved performance and stability, an investigation into the crystallization of perovskite on diverse interfaces is undertaken. Primarily, the bisFAPA exhibits significantly enhanced solubility in 1-butanol (**Figure S8**), due to improved hydrophilicity of the molecule by doubling the substitution of the fullerene moiety with phosphonic acids. The devices obtained upon optimizing the scalable deposition of FAPA by blade coating show comparable performance compared to those prepared by spin-coating the ETL with FAPA (**Figure S9**). Secondly, as illustrated in **Figure S10**, the increased water contact angle, transitioning from 52º to 69º on SnO₂ modified with monoFAPA and bisFAPA, respectively, signifies the reducing surface energy upon modification of the SnO₂ surface with double phosphonic acid substituted fullerene compared to monoFAPA. The heightened hydrophobicity might result from a denser bisFAPA layer on SnO₂. The reduced surface energy leads to improved crystal grains of the perovskite grown on top of the SnO₂ layer, in agreement with previous reports.[25] The ultimate perovskite morphology on various interfaces is revealed in the cross-sectional scanning electron microscope (SEM) image in **Figure 3a-c**. The contact on bare SnO₂ exhibits small crystals with a random distribution, resulting in an undesired heterojunction. In contrast, a noticeable enlargement of crystal size is observed on the FAPA-modified interface, particularly on the bisFAPA, where the crystal grains are most prominently expanded. Moreover, the discernible differences are accentuated in the x-ray diffraction (XRD) pattern in **Figure 3d-f**. The heightened intensity of the (110) crystal plane, particularly evident with bisFAPA modification, signifies an enhanced order of crystallinity and an increase in crystal size. Furthermore, the full-width at half maximum (FWHM) of the (110) and (220) peaks in the reference decreased by 63% and 41% respectively with bisFAPA modification. This reduction further indicates an enhanced crystalline quality and improved structural characteristics. The outcomes consistently affirm the heightened crystallinity and improved morphology of perovskite on enhanced interfaces and the elevated quality of perovskite, particularly at the modified interface, elucidates the observed commendable stability in both illumination and mechanical aspects. As depicted in **Figure 3g-i**, the perovskite film exhibits barely no cracks on FAPA modified interfaces, even after enduring 3000 bending cycles.

The resulting response can explain why the device was able to maintain 95% of its initial PCE in the bending test.

**Effects of FAPA on SnO₂ electronic structure**

To understand the role of FAPA in device improvement, we turn to investigate how it chemically and electronically affects the SnO₂ interface. X-ray photoelectron spectroscopy (XPS) measurements prove the presence of FAPA on the SnO₂ surface through the conspicuous phosphorus signals in the XPS spectrum (**Figure S11**). **Figure 4a** displays the narrow XPS scans of the Sn $3d$ core level measured on the SnO₂, SnO₂/monoFAPA, and SnO₂/bisFAPA surfaces. The intensity of the Sn $3d$ peak decreases significantly upon modification of the SnO₂ layer by monoFAPA and is further reduced in the case of bisFAPA, suggesting that bisFAPA provides a much denser fullerene layer than monoFAPA, which is likely due to the enhanced chemical interaction with the SnO₂ surface provided by two phosphonic groups per molecule. A rough estimate of the effective film thicknesses of FAPA layers on top of SnO₂ is obtained from the drop of the Sn signal in the XPS spectrum, yields values of approximately 1-3 nm, by using the Beer-Lambert law: $x = -(1/\alpha) \times \ln(I/I_0)$, where $I$ is the measured intensity of the Sn signal, $I_0$ is the initial intensity of the Sn signal (reference), α is the attenuation factor and x is the effective film thickness of the FAPA layer.[26-28]

The XPS O $1s$ spectra in **Figure 4b** demonstrate a substantial chemical reaction between phosphonic acids and the hydroxyl groups (-OH) at the SnO₂ surface. Capping the SnO₂ surface with FAPA thin films results in a marked reduction in the intensity of -OH groups due to heterocondensation with the phosphonic acid group by formation of Sn-O-P bonds,[29-31] while part of the observed effect may also be attributed to the increased escape depth of electrons that originate from the -OH groups, by the presence of the FAPA layer.[32] In comparison to monoFAPA, bisFAPA exhibits a more pronounced O=P peak, at an approximately unchanged O-C signal, suggesting that the surface density of fullerene moieties is similar for both FAPA derivatives.[33] Consequently, the surface density of phosphonic acid groups on bisFAPA-modified SnO₂ is about twice that of monoFAPA, which leads to an overall higher surface coverage in the case of the latter. **Figure 4c** illustrates the O=P binding of bisFAPA to the SnO₂ surface, which leads to the formation of a

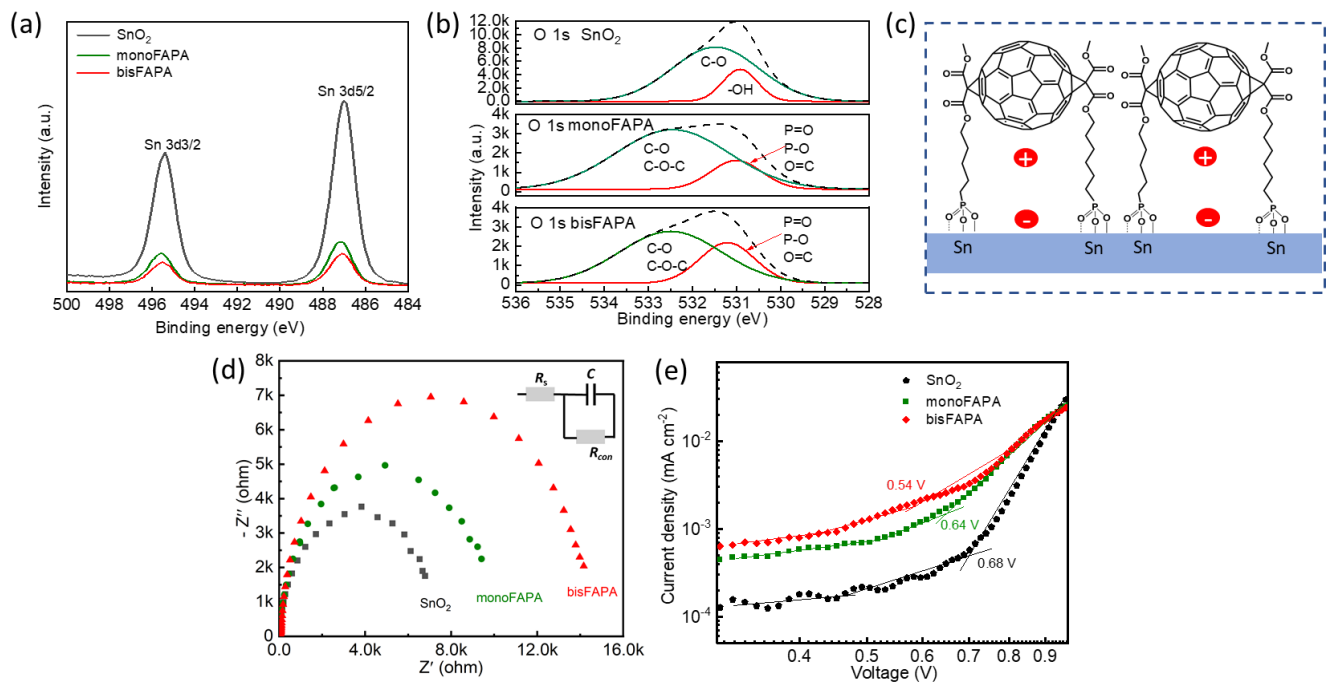

Figure 4 (a) XPS peaks of Sn 3*d*. (b) O 1s XPS on thin films of SnO₂, SnO₂ modified with monoFAPA and bisFAPA, respectively. The dashed lines represent the raw data, while the solid lines represent the peak signal that has been fitted. The C-O peak on SnO₂ surface likely stems from the ligands contained in the SnO₂ nanoparticle ink. (c) Schematic illustration of bisFAPA bonding to the SnO₂ surface. (d) Nyquist plots of perovskite solar cells prepared on bare SnO₂ and SnO₂ interfaces with FAPA modification. The inset shows the corresponding equivalent circuit. (e) SCLC of electron-only devices with SnO₂ interfaces either unmodified or modified with two different FAPA layers.

dipole layer, with the dipole moment pointing towards the SnO₂ surface.[34] Due to the higher number of phosphonic acid groups per fullerene molecule, the dipole moment is expected to be higher for bisFAPA layers than for those formed from monoFAPA.

The modification of SnO₂ by FAPA is further supported by impedance measurements of the f-PSCs. The Nyquist plots shown in **Figure 4d** are obtained by oscillating the frequency from 0.1 Hz to 1 MHz in the dark at open circuit conditions. The data are interpreted in terms of an equivalent circuit which assigns the semicircle to the SnO₂/perovskite interface, represented by a contact resistance and a capacitor (inset of **Figure 4d**).[35, 36] The contact resistance increases from $R_{con}$ = 6.9 kΩ for the unmodified SnO₂ to $R_{con}$ = 9.8 kΩ for the monoFAPA-modified ETL and further to $R_{con}$ = 14.4 kΩ for bisFAPA. The observed rise in contact resistance is attributed to a denser FAPA layer, providing improved surface coverage and acting as an electric barrier against the recombination of electrons and holes.[37] Furthermore, it is interesting to note that this increase of contact resistance at the SnO₂/perovskite interface shows the same trend as the suppression of hysteresis. This effect is at least partially attributable to reduced ionic defect density in the bulk of perovskite due to improved crystal growth on top of the FAPA layers, as demonstrated in the previous section.

To provide a measure for the effect of SnO₂ modification by FAPA on perovskite crystal quality, we have estimated the charge carrier trap density in modified and unmodified devices, by employing the space-charge-limited current (SCLC) technique using single charge-transporting structures. For this purpose, ETL-only devices composed of a PET/IMI/SnO₂ layer, with and without FAPA, followed by a perovskite layer, PCBM layer, and a carbon electrode were fabricated. **Figure 4e** displays a logarithmic presentation of the dark *I-V* curves, contrasting devices with and without fullerene derivatives at the interface. In the low bias voltage regime, a linear response, indicative of ohmic behaviour is observed. As bias voltage increases, a significant rise in current injection occurs in the intermediate region. It is obvious that the introduction of monoFAPA and bisFAPA layers onto the SnO₂ surface leads to a reduction in trap-filling voltage ($V_{TFL}$). Consequently, the trap density ($N_t$) within the devices is derived from the following relationship[38]: $N_t = \frac{2\varepsilon_0\varepsilon_P V_{TFL}}{eL_D}$, where $\varepsilon_0$ is the vacuum dielectric constant, $e$ is the

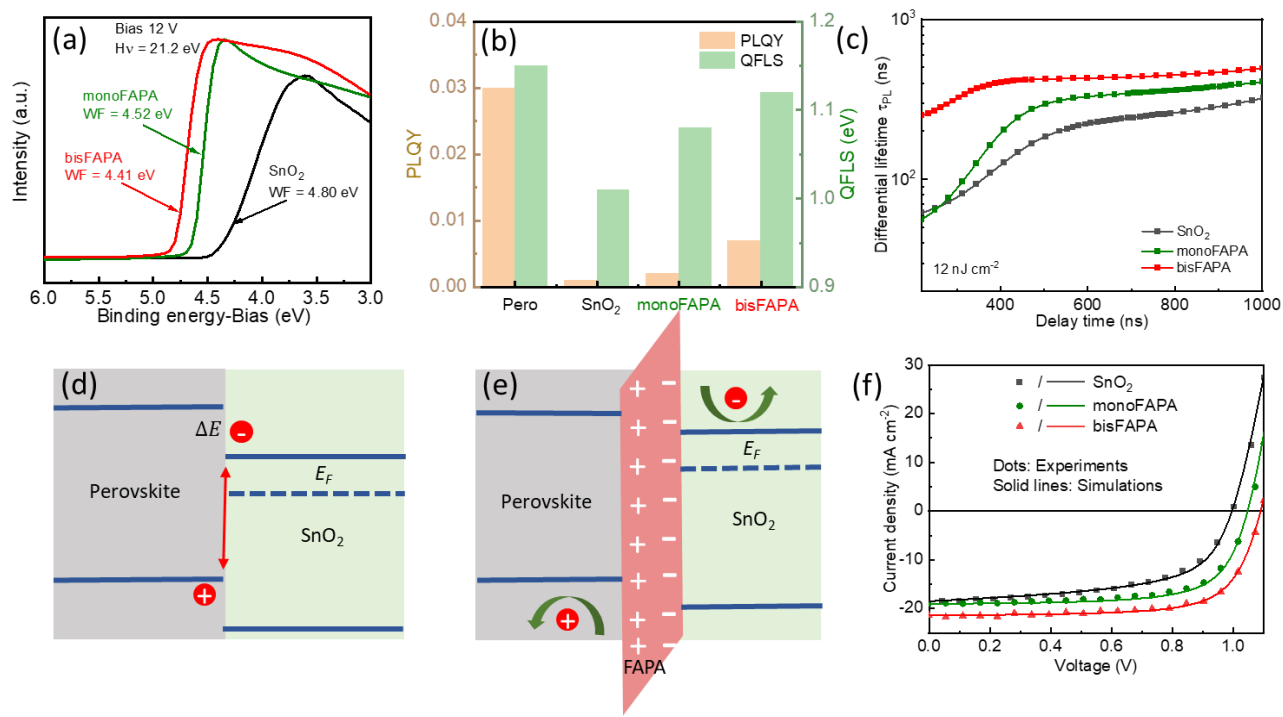

Figure 5 (a) UPS spectra showing the secondary cut-off energies and the resulting Fermi levels of the SnO₂ surfaces with and without FAPA modification. (b) PLQY and QFLS of the perovskite on bare SnO₂, SnO₂/monoFAPA and SnO₂/bisFAPA. (c) Differential transient lifetimes of perovskite on bare SnO₂, SnO₂/monoFAPA and SnO₂/bisFAPA at a laser fluence of 12 nJ cm⁻². (d) Schematic drawing of the energy band diagram at the interface between perovskite and the unmodified SnO₂ interface. (e) Schematic drawing of the energy band diagram of the interface between perovskite and FAPA-modified SnO₂. (f) Simulated *JV* curves match well with the experimental results.

electron charge, $\varepsilon_p$ is the perovskite dielectric constant, and $L_D$ is the perovskite layer thickness. The trap density in the bulk perovskites decreases from 9.4×10¹⁵ cm⁻³ for unmodified SnO₂ to 8.9×10¹⁵ cm⁻³ and 7.4×10¹⁵ cm⁻³ for monoFAPA and bisFAPA, respectively, confirming an improved quality of perovskite films grown on FAPA-modified SnO₂.

To understand how the FAPA modification and the resulting energy level alignment at the SnO₂/perovskite interface affect device performance, we turn to ultraviolet photoelectron spectroscopy (UPS) measurements. The presence of the dipole layer formed by FAPA generates a localized electric field at the SnO₂/perovskite interface which upshifts the Fermi level ($E_F$) of SnO₂, which is shown in **Figure 5a**. The $E_F$ values are determined from the positions of the secondary electron cut-off energies ($E_{cut-off}$) by the equation $E_F = h\nu$-Bias- $E_{cut-off}$. The $E_F$ values of SnO₂ thus calculated decrease from -4.80 eV to -4.52 eV upon modification of SnO₂ with monoFAPA, and further drop to -4.41 eV for bisFAPA, in accordance with the expected stronger interface dipole created by the latter. These values are in agreement with the work functions of the corresponding samples determined by Kelvin probe measurements (**Figure S12**).

The impact of energy level alignment on interfacial charge recombination is quantified by measurements of photoluminescence quantum yield (PLQY) and time-resolved PL decay traces. **Figure 5b** shows the measured PLQY of perovskite coated on bare SnO₂ and SnO₂ modified with interface dipole layer. Notably, the introduction of FAPA dipole layers on top of SnO₂ leads to an enhancement in PLQY, by a factor of 2.1 for monoFAPA and by a factor of 7.3 for bisFAPA, in accordance with the PL decay times extracted from **Figure S13.** The enhanced PL lifetimes are clearly demonstrated by the differential lifetime plot shown in **Figure 5c**. A higher differential lifetime corresponds to lower surface recombination at low injection due to less charge carrier accumulation at interfaces (details of the calculations are provided in SI).[39] The corresponding minority carrier lifetimes obtained by fitting bi-exponential decay functions to the decay curves are shown in **Table S1**.

This increase in PLQY and PL lifetime is explained by the strongly reduced interfacial non-radiative recombination upon the introduction of the FAPA interface dipole layer as illustrated in

**Figure 5d-e.** A significant energetic offset between the ETL and perovskite promotes the recombination of electrons and holes through surface traps, leading to pronounced interfacial recombination (**Figure 5d**).[40] Introducing an interface dipole layer by modification of SnO₂ with FAPA elevates the Fermi level of the ETL, leading to a smaller offset of the conduction band minima at the SnO₂/perovskite interface (**Figure 5e**) and thus to reduced band bending at this interface under illumination, which in turn serves to efficiently release photogenerated charge carriers from the interface.[40] The consequent diminution of interface charge carrier density effectively mitigates interface recombination. As in our system interface recombination is the dominating process of charge carrier loss, the effect of reinjection of extracted minority carriers is negligible, which overall leads to an increase of minority carrier lifetime. Consequently, the suppression of interface recombination, enabled by the reduced energetic offset, translates into an overall improvement in photovoltaic performance.

The effect of reduced interface recombination on quasi-Fermi level splitting (QFLS) is calculated according to $QFLS = V_{oc}^{rad}(E_g) + kTln(PLQY)$, where the $V_{oc}^{rad}$ is radiative Shockley-Queisser limit is related to the optical bandgap of perovskite.[41] The introduction of monoFAPA and bisFAPA thus results in a significant improvement of the external QFLS of the perovskite on SnO₂, with values increasing from 1.01 eV for bare SnO₂ to 1.08 eV and 1.12 eV, respectively. This result is in good agreement with the $V_{OC}$ values of our devices represented above.

To further evaluate how energy level alignment affects device performance, we turn to drift-diffusion simulations of f-PSCs using the open-source package SIMsalabim to perform the simulations and BOAR to perform the $JV$ fitting using Bayesian optimization.[42-46] The primary parameter under investigation is the layer stack of SnO₂ with FAPA modification. Other parameters, such as the energetics of SnO₂ and the interfacial trap density, are incorporated into the simulation based on experimental results and documented data (**Table S2**). The simulated $JV$ curves under 1 sun illumination, **Figure 5f**, show a good match with the experimental $JV$ curves. The device with the interface dipole modified SnO₂ outperforms the device with only bare SnO₂, due to increased $V_{OC}$ and $FF$. The simulation results confirm the effectiveness of the introduced interface dipole on SnO₂ in reducing interfacial recombination between ETL and perovskite.

We can more quantitively describe the contributions from energetic offset and interfacial traps to the overall interfacial recombination rate ( $w_{int}$ ), through the equation $w_{int} = C_{min}N_{t,int}\frac{N_{c,ETL}}{N_{c,Pero}}\exp(\frac{\Delta E_c}{K_bT})$,[47] where $C_{min}$ is minority carrier capture coefficient, and $N_{t,int}$ is the interfacial trap density that determined from the trPL lifetime, as detailed in the SI. $N_{c,ETL}$ and $N_{c,Pero}$ are the effective densities of states of ETL and perovskite, respectively. $\Delta E_c$ is the energetic offset between conduction band (CB) of perovskite and CB of ETL and $k_bT$ is thermal energy. From bare SnO₂ to monoFAPA and bisFAPA modification, the decreased energetic offset $\Delta E_c$ reduces $w_{int}$ from $7.53 \times 10^4$ m⁻¹s⁻¹ to $7.99 \times 10^3$ m⁻¹s⁻¹ by monoFAPA and further to $3.04 \times 10^1$ m⁻¹s⁻¹ by bisFAPA (**Table S2**). The trend observed in the simulation, where a significant reduction in interfacial recombination rates is achieved by lowering the energetic barrier between ETL and the perovskite, aligns with the trend observed in our experiments through the decrease in the WF of the ETL as measured by UPS. This supports our hypothesis that the introduction of a thin dipole layer in the ETL alters the electronic energy band structure at the ETL interface, thus substantially reducing non-radiative recombination at the buried interface. Consequently, this enhancement in optoelectronic performance is evident in the improved device characteristics.

## Upscaling solar module fabrication

Having demonstrated the beneficial impact of FAPA interlayers on solar cell performance, in the last section, we turn to consider upscaling the production of bisFAPA-modified f-PSMs.

**Figure 6a** shows a photograph of the f-PSM consisting of seven solar cells connected in series, with an active area of 2.41 cm² for each individual cell and a total aperture area of 20.25 cm² on flexible substrate. As illustrated in **Figure 6b**, the cell interconnection within the module requires three parallel scribes designated as P1, P2, and P3. To electrically separate the IMI bottom electrode into the desired number of individual cells, P1 lines are scribed on flexible PET/IMI sheets using a femtosecond laser with a fluence of 400 $\mu J$ cm⁻². The spikes that are typically formed by laser ablation of IMI films are removed afterwards by R2R CO₂ snow jetting.[48] After coating the functional layers on the P1 patterned IMI substrates, P2 lines are laser-ablated, which serves as a prerequisite for the electrical interconnection of adjacent cells upon coating the top carbon electrode.

To underline the significance of this work, we summarize in **Figure 6f** the recent advancements in the efficiency of f-PSCs. Details of device parameters are shown in **Table S3**. The majority of the state-of-art f-PSCs reported over the past eight years were fabricated using the spin-coating technique and evaporated metal as top electrode. Among the fully printed f-PSCs with

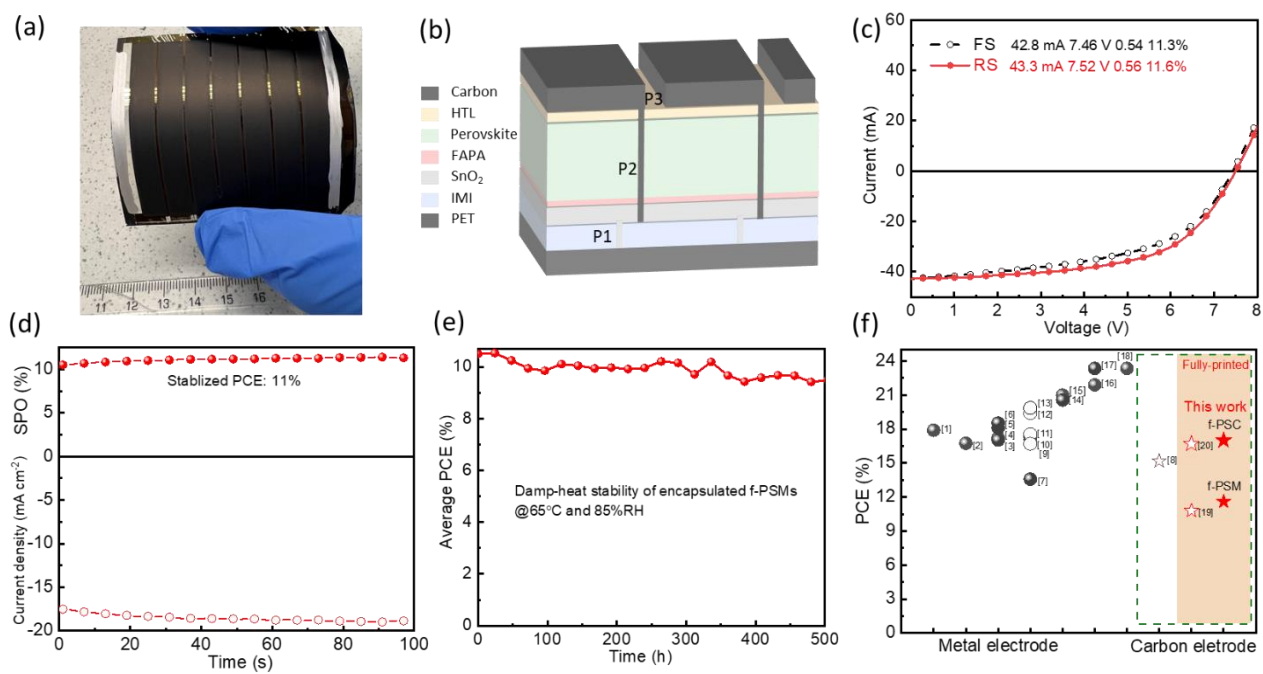

Figure 6 (a) Photograph of f-PSM on an area of 25 cm² plastic substrate. (b) Cross-sectional schematic drawing showing how individual cells are connected in series to form the thin-film solar modules. (c) I-V curves of the champion flexible module measured with FS and RS. (d) SPO of the f-PSM with bisFAPA modification. (e) Damp-heat stability test for encapsulated f-PSMs in the best condition with bisFAPA modification. (f) Overview of PCE of f-PSCs deposited by spin coating (solid black circle) and scalable techniques (open black circle) with evaporated metal electrodes and fully printed f-PSCs with carbon electrode (light red area). The numbers behind the dots refer to the corresponding references which can be found in the SI.

With the dedicated modification of the P1, P2 and P3 line fabrication (details in SI), the resulting champion module manufactured with the bisFAPA modified SnO₂ layer shows a PCE of 11.6% ($V_{OC}$ = 7.52 V, $I_{SC}$ = 43.3 mA and $FF$ = 56%), with minimized hysteresis, **Figure 6c**. The SPO, by holding the module at the voltage of maximum power point and measuring the photocurrent, is 11.0% as shown in **Figure 6d**. Finally, we conducted a damp-heat test on the encapsulated f-PSMs. The modules are encapsulated inside a barrier foil, carried out under inert conditions using rim type encapsulation as depicted in **Figure S14**. Subsequently, the encapsulated f-PSMs are subjected to accelerated lifetime testing under damp heat conditions using ISOS-D3 (65 °C and 85% RH) as the test standard (**Figure S15**).[49] As depicted in **Figure 6e**, these modules maintain approximately 90% of their initial PCE after a 500 h test period.

carbon electrode reported so far, this work presents the highest PCE. Most importantly, this work represents the highest PCE reported on a fully printed f-PSM with metal-free electrode. A fully printed perovskite module on flexible substrate with a PCE of 11.6% represents a significant advancement in the field of printed perovskite PV.

## Conclusions

This study presents the scalable fabrication of efficient, fully printed f-PSMs comprising printed carbon top electrodes, through careful optimization of interfacial modifier, layer coating conditions and laser scribing parameters. Special attention has been paid to the perovskite/ETL interface, at which non-radiative recombination loss has been substantially reduced by the construction of a continuous and dense FAPA thin layer on top of the SnO₂ layer, leading to remarkable improvement in $V_{OC}$ and $FF$. We find that an interfacial dipole layer is formed, through strong chemical bonding between phosphonic acid and the hydroxyl groups of $SnO_2$, which reduces the WF of $SnO_2$ and thus the energetic offset with respect to the perovskite layer. Through the refinement of all interfaces via scalable coating methods, PCE values of 17.0% and 11.6% for f-PSCs and f-PSMs have been achieved, respectively, with long-term mechanical durability and consistent operational stability. These findings, particularly the production of large-scale flexible modules, achieve substantial progress toward the eventual commercialization of printable perovskite PV.

## Author Contributions

L.D. performed the experiments and wrote the manuscript. S.Q. was involved in the discussion of the manuscript. J.G.C. made the module layout. M.W. printed the ETL layer on flexible substrate. O.K. measured the XPS spectrum. H.P. measured the UPS samples. S.F encapsulated the flexible modules. D.J. and C.L. measured and analysed the PL data. K.Z. measured infrared absorption. V.L.C. did the simulation part. F.U.K. and F.Y. modified the manuscript. C.A. ordered the materials. C.J.B. reviewed the manuscript. H.-J.E. supervised the work and the writing of the manuscript. All the co-authors discussed the results and commented on the manuscript.

## Conflicts of interest

The authors declare no conflict of interest.

## Acknowledgements

The Solar Factory of the Future (SFF) as part of the Energy Campus Nürnberg (EnCN) is acknowledged, which is supported by the Bavarian State Government (FKZ 20.2-3410.5-4-5). Part of this work has been supported by the Helmholtz Association in the framework of the innovation platform "Solar TAP". L.D. and T.D. acknowledges the financial support from Deutsche Forschungsgemeinschaft (DFG) via the Perovskite SPP2196 program (project no.506698391). L.D. gratefully acknowledges funding of the Erlangen Graduate School in Advanced Optical Technologies (SAOT) by the Bavarian State Ministry for Science and Art. S.Q. and C.L. are grateful for the support from the China Scholarship Council (CSC). J.G.C. gratefully acknowledges the Deutscher Akademischer Austauschdienst (DAAD) for the granting of a doctoral scholarship. F.U.K. thanks the Jardine Foundation and Cambridge Trust for a doctoral scholarship. F. Y. acknowledges the financial support from the National Natural Science Foundation of China (Grant Nos. 52102287), Natural Science Foundation of Jiangsu Province (Grant No. BK20210731), the Natural Science Foundation of the Jiangsu Higher Education Institutions of China (21KJD150003).

## References

1  Z. Li, C. Jia, Z. Wan, J. Xue, J. Cao, M. Zhang, C. Li, J. Shen, C. Zhang and Z. Li, *Nat. Commun*, 2023, **14**, 6451.

2  L. Xie, S. Du, J. Li, C. Liu, Z. Pu, X. Tong, J. Liu, Y. Wang, Y. Meng and M. Yang, *Energy Environ. Sci*, 2023, **16**, 5423-5433.

3  Y. Wu, G. Xu, J. Xi, Y. Shen, X. Wu, X. Tang, J. Ding, H. Yang, Q. Cheng and Z. Chen, *Joule*, 2023, **7**, 398-415.

4  B. Dou, J. B. Whitaker, K. Bruening, D. T. Moore, L. M. Wheeler, J. Ryter, N. J. Breslin, J. J. Berry, S. M. Garner and F. S. Barnes, *Acs Energy Lett*, 2018, **3**, 2558-2565.

5  D. Beynon, E. Parvazian, K. Hooper, J. McGettrick, R. Patidar, T. Dunlop, Z. Wei, P. Davies, R. Garcia - Rodriguez and M. Carnie, *Adv. Mater* 2023, **35**, 2208561.

6  T. Bu, J. Li, F. Zheng, W. Chen, X. Wen, Z. Ku, Y. Peng, J. Zhong, Y.-B. Cheng and F. Huang, *Nat. commun*, 2018, **9**, 4609.

7  Y. Y. Kim, T.-Y. Yang, R. Suhonen, A. Kemppainen, K. Hwang, N. J. Jeon and J. Seo, *Nat. Commun*, 2020, **11**, 5146.

8  H. C. Weerasinghe, N. Macadam and J.-E. Kim, *Nature Communications*, 2024, **15**, 1656.

9  F. Jafarzadeh, L. A. Castriotta, F. De Rossi, J. Ali, F. Di Giacomo, A. Di Carlo, F. Matteocci and F. Brunetti, *Sustain. Energy Fuels*, 2023, **7**, 2219-2228.

10  L. J. Sutherland, D. Vak, M. Gao, T. A. N. Peiris, J. Jasieniak, G. P. Simon and H. Weerasinghe, *Adv. Energy Mater*, 2022, **12**, 2202142.

11  E. H. Jung, N. J. Jeon, E. Y. Park, C. S. Moon, T. J. Shin, T. Y. Yang, J. H. Noh and J. Seo, *Nature*, 2019, **567**, 511-515.

12  F. Yang and L. Dong, *Adv. Energy Mater*, 2021, **11**.

13  F. Yang, L. Dong and H. J. Egelhaaf, *Adv. Energy Mater*, 2020, **10**.

14  T. Du, S. Qiu, X. Zhou and H.-J. Egelhaaf, *Joule*, 2023, **7**, 1920-1937.

15  F. Giordano, A. Abate, J. P. Correa Baena, M. Saliba, T. Matsui, S. H. Im, S. M. Zakeeruddin, M. K. Nazeeruddin, A. Hagfeldt and M. Graetzel, *Nat. Commun*, 2016, **7**, 10379.

16  Z.-L. Tseng, C.-H. Chiang and C.-G. Wu, *Sci. Rep*, 2015, **5**, 13211.

17  S. Y. Park and K. Zhu, *Adv.Mater*, 2022, **34**, 2110438.

18  Z. Li, T. R. Klein, D. H. Kim, M. Yang, J. J. Berry, M. F. Van Hest and K. Zhu, *Nat. Rev. Mater*, 2018, **3**, 1-20.

19  M. Valles-Pelarda, B. C. Hames, I. Garcia-Benito, O. Almora, A. Molina-Ontoria, R. S. Sanchez, G. Garcia-Belmonte, N. Martin and I. Mora-Sero, *J Phys Chem Lett*, 2016, **7**, 4622-4628.

20  D. Garcia Romero, L. Di Mario, F. Yan, C. M. Ibarra‐Barreno, S. Mutalik, L. Protesescu, P. Rudolf and M. A. Loi, *Adv. Energy Mater*, 2023, 2307958.

21  G. Tumen-Ulzii, T. Matsushima, D. Klotz, M. R. Leyden, P. Wang, C. Qin, J.-W. Lee, S.-J. Lee, Y. Yang and C. Adachi, *Commun. Mater*, 2020, **1**, 31.

22  Y. Hou, S. Scheiner, X. Tang, N. Gasparini, M. Richter, N. Li, P. Schweizer, S. Chen, H. Chen, C. O. R. Quiroz, X. Du, G. J. Matt, A. Osvet, E. Spiecker, R. H. Fink, A. Hirsch, M. Halik and C. J. Brabec, *Adv. Mater. interfaces*, 2017, **4**.

23  Y. Hou, X. Du, S. Scheiner, D. P. McMeekin, Z. Wang, N. Li, M. S. Killian, H. Chen, M. Richter and I. Levchuk, *Science*, 2017, **358**, 1192-1197.

24  B. Chen, M. Yang and S. Priya, *The journal of physical chemistry letters*, 2016, **7**, 905-917.

25  D. Liu, Z. Shao, J. Gui, M. Chen, M. Liu, G. Cui, S. Pang and Y. J. C. c. Zhou, 2019, **55**, 11059-11062.

26  D. Y. Zemlyanov, M. Jespersen and D. N. Zakharov, *Nanotechnology*, 2018, **29**, 115705.

27  P. J. S. Cumpson, *interfaces thin films*, 2000, **29**, 403-406.

28  F. A. Stevie and C. L. Donley, *Journal of Vacuum Science Technology A*, 2020, **38**.

29  P. H. Mutin, G. Guerrero and A. Vioux, *J Mater Chem A*, 2005, **15**, 3761-3768.

30  S. A. Paniagua, P. J. Hotchkiss and S. C. Jones, *The Journal of Physical Chemistry C*, 2008, **112**, 7809-7817.

31  S. Pawsey, K. Yach and L. Reven, *Langmuir*, 2002, **18**, 5205-5212.

32  F. Ahimou, C. J. Boonaert, Y. Adriaensen, P. Jacques, P. Thonart, M. Paquot and P. G. Rouxhet, *J Colloid Interf Sci*, 2007, **309**, 49-55.

33  E. Yalcin, M. Can, C. Rodriguez-Seco, E. Aktas, R. Pudi, W. Cambarau, S. Demic, E. J. E. Palomares and E. Science, *Energy. Environ. Sci*, 2019, **12**, 230-237.

34  R. Wick-Joliat, T. Musso, R. R. Prabhakar, J. Löckinger, S. Siol, W. Cui, L. Sévery, T. Moehl, J. Suh and J. E. Hutter, *Energy Environ Sci*, 2019, **12**, 1901-1909.

35  F. Ebadi, N. Taghavinia, R. Mohammadpour, A. Hagfeldt and W. Tress, *Nat. Commun*, 2019, **10**, 1574.

36  A. Guerrero, J. Bisquert and G. Garcia-Belmonte, *Chem Rev*, 2021, **121**, 14430-14484.

37  E. Von Hauff and D. Klotz, *Journal of Materials Chemistry C*, 2022, **10**, 742-761.

38  V. M. Le Corre, E. A. Duijnstee, O. El Tambouli, J. M. Ball, H. J. Snaith, J. Lim and L. J. A. Koster, *Acs Energy Lett*, 2021, **6**, 1087-1094.

39  L. Kruckemeier, Z. Liu and T. Kirchartz, *Adv Mater*, 2023, **35**, e2300872.

40  L. Krückemeier, B. Krogmeier, Z. Liu, U. Rau and T. Kirchartz, *Adv. Energy. Mater*, 2021, **11**.

41  M. Stolterfoht, P. Caprioglio, C. M. Wolff, J. A. Márquez, J. Nordmann, S. Zhang, D. Rothhardt, U. Hörmann, Y. Amir, A. Redinger, L. Kegelmann, F. Zu, S. Albrecht, N. Koch, T. Kirchartz, M. Saliba, T. Unold and D. Neher, *Energy. Environ. Sci*, 2019, **12**, 2778-2788.

42  https://github.com/kostergroup/SIMsalabim).

43  https://github.com/i-MEET/boar).

44  M. Wagner, A. Distler, V. M. L. Corre, S. Zapf, B. Baydar, H.-D. Schmidt, M. Heyder, K. Forberich, L. Lüer and C. Brabec, *Energy. Environ. Sci*, 2023.

45  C. Liu, L. Lüer, V. M. L. Corre, K. Forberich, P. Weitz, T. Heumüller, X. Du, J. Wortmann, J. Zhang and J. Wagner, *Adv.Mater*, 2023, 2300259.

46  M. Koopmans, V. M. Le Corre and L. J. A. Koster, *J. Open Source Softw*, 2022, **7**, 3727.

47  M. Koopmans and L. J. A. Koster, *Sol Rrl*, 2022, **6**.

48  M. Wagner, A. Distler, H.-D. Schmidt, A. Classen, T. Stubhan, M. Koegl, J. Illg, C. Brabec and H.-J. Egelhaaf, *Flex. Print. Electron*, 2023, **8**, 015007.

49  M. V. Khenkin, E. A. Katz, A. Abate, G. Bardizza, J. J. Berry, C. Brabec, F. Brunetti, V. Bulović and Q. Burlingame, *Nat. Energy*, 2020, **5**, 35-49.